\documentclass[11pt]{article}
\usepackage{amssymb}
\usepackage{amsmath,amsthm,amsfonts}  
\usepackage{graphicx}
\usepackage[applemac]{inputenc}  

\newcommand{\ket}[1]{|\kern.3ex#1\kern.3ex\rangle}
\newcommand{\bra}[1]{\langle\kern.3ex #1 \kern.3ex|}
\newcommand{\scalar}[2]{\langle\kern.3ex #1 \kern.3ex|\kern.3ex#2\kern.3ex\rangle}
\title{Wigner distribution}

\begin{document}

\title{About Statistical Questions Involved in the Data Analysis of the OPERA Experiment}
\author{Hervé BERGERON$^{1}$ \thanks{herve.bergeron@u-psud.fr} \\
 $^1$ \textit{Universit\'e Paris-Sud, ISMO, UMR 8214 du CNRS}, \\ B\^at. 351 F-91405 Orsay, France}

\maketitle
{\abstract{ The authors of the OPERA experiment [ArXiv: 1109.4897] claim that ``{\it the measurement indicates an early arrival time of CNGS muon neutrinos with respect to the one computed assuming the speed of light in vacuum}''. In this note we analyze the statistical aspects of the experimental results presented in [ArXiv: 1109.4897], assuming that no hidden experimental bias exists. Due to statistical constraints, we show (through two different methods) that the experimental data presented in [ArXiv: 1109.4897] do not permit to conclude unambiguously with the existence of a superluminal behavior of neutrinos. The problem lies essentially in the interpretation of the data and not in their veracity.}}

\bigskip
\tableofcontents
\newpage

\section{Introduction}
The OPERA experiment attempts to measure the velocity of neutrinos using a {\it one way} method,  the starting point being at the CERN and the end point being at the Laboratorio Nazionale del Gran Sasso (LNGS). This method needs an accurate knowledge of the distance baseline and a detailed budgeting of all delays with a very accurate synchronization of clocks at both ends of the baseline. 
The stated result, after correction of all systematic bias, is that the Time Of Flight (TOF) of the neutrinos is shorter by $60 \pm 6.9 \, {\rm (stat.)} \pm 7.4 \, {\rm (sys.) \, ns}$ than that expected for a travel at the speed of light. \\
A large number of papers dedicated to this article are already present on {\it ArXiv} \cite{Cacciapaglia,Amelino,Auterio,Ciborowski,Alexandre,Klinkhamer,Giudice,Dvali,Gubser,Nikolic,Naumov,Contaldi, Sanctis}, this list representing only few examples. They can be roughly divided into three categories. The papers that try to justify or invalidate the results of OPERA  on a pure theoretical level, the comments on the latter that question the theoretical arguments developed,  and the papers that question the experimental methodology of OPERA and the possible experimental biases (as clock synchronization \cite{Contaldi}). Here we address the question of whether the stated result is consistent from a purely data analysis point of view (only using the informations given in \cite{Adam}). Moreover we assume that the experimental results presented in \cite{Adam} do not contain any hidden systematic bias.\\
In order to develop our analysis, we need first to recall briefly the experimental results obtained by OPERA and how these data have been treated to lead to the claim of a superluminal behavior of neutrinos. Then we apply our data analysis to address the question of whether a superluminal effect can be deduced from the data.

\section{A simplified presentation of OPERA results and their data analysis (with comments)}
For more details see \cite{Adam}.
\subsection{The experiment and its framework}
The physical phenomenon  observed by OPERA can be roughly seen as being the propagation of an energetic pulse transmitted by different particles over time. At the beginning the energetic pulse is a proton one (selected in energy) that is partially converted (by collision on a target) into a meson one ($K$ or $\pi$) which itself decays into a neutrino one. These steps occur at CERN. Then the neutrino pulse travels about $730 {\rm \,km}$ and is (partially) detected at LNGS. Moreover, because of the domain of energy involved, mesons and neutrinos can be assumed to travel at the speed of light (with a high precision).\\
The initial proton pulse is characterized on the target by its instantaneous courant represented by the {\it waveform} $w(t)$. The neutrinos detected at LNGS allow to rebuild in principle  the instantaneous neutrino rate $\phi_\nu(t)$ (see the section below for the practical problems encountered). Then the simple relation expected is
\begin{equation}
\phi_\nu(t)=K \, w(t-\Delta) \; {\rm with} \; \Delta=L/c,
\end{equation}
where $K$ is some constant of proportionality and $L$ is the distance traveled by the particles.\\
In fact the situation is more complicated due to the low efficiency of neutrinos detection. Then an average between different experiments (different pulses) is needed, leading to the averaged quantities $\bar{w}(t)$ and $\bar{\phi}_\nu(t)$ that are expected to verify the previous relation
\begin{equation}
\label{eqn:propor}
\bar{\phi}_\nu(t)=K \, \bar{w}(t-\Delta).
\end{equation}
The authors of \cite{Adam} claim that this relation is not verified, but the following delayed one
\begin{equation}
\label{eqn:propordelay}
\bar{\phi}_\nu(t)=K \, \bar{w}(t-\Delta-\tau).
\end{equation}
is verified (after correction of all systematic bias) with the negative delay
\begin{equation}
\tau=-60 \pm 6.9 \, {\rm (stat.)} \pm 7.4 \, {\rm (sys.)},
\end{equation}
stating an early arrival of neutrinos and then a superluminal neutrino effect. So their claim is divided into two different assertions:
\begin{itemize}
\item[(a)] The expected delayed proportionality rule \eqref{eqn:propor} is false,
\item[(b)] The new delayed relation \eqref{eqn:propordelay} is true.
\end{itemize}
Since there is no direct logical inference between these two assertions, let us remark on a pure logical level that (a) can be true, (b) being false. In other words it is possible that the relation between $\bar{w}(t)$ and $\bar{\phi}_\nu(t)$ is not at all a simple delayed rule of proportionality.

\subsection{The data}
\subsubsection{The averaged proton waveforms}
\begin{figure}[!ht]
\resizebox{\hsize}{!}{\includegraphics{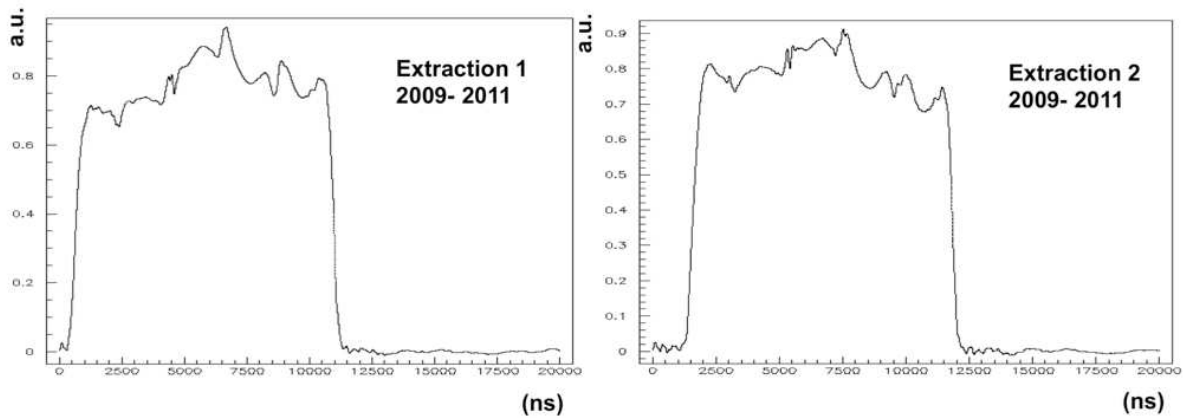}}
\caption{The averaged waveforms $\bar{w}_1(t)$ and $\bar{w}_2(t)$ given in \cite{Adam} (fig.9).} 
\label{figure1}
\end{figure}

The averaged waveform $\bar{w}(t)$ is given in \cite{Adam} (fig.9) which is here reproduced in fig.\ref{figure1}. In fact two different waveforms 
$\bar{w}_1(t)$ and $\bar{w}_2(t)$ are given corresponding to two {\it extractions} that is two successive proton pulses. Each pulse $\bar{w}_i(t)$ has an approximate rectangular shape and a duration of about $10\; \mu{\rm s}$ (for a stated superluminal effect of about $60 \, {\rm ns}$).

\subsubsection{The detected rate of neutrinos}
\begin{figure}[!ht]
\includegraphics[scale=0.8]{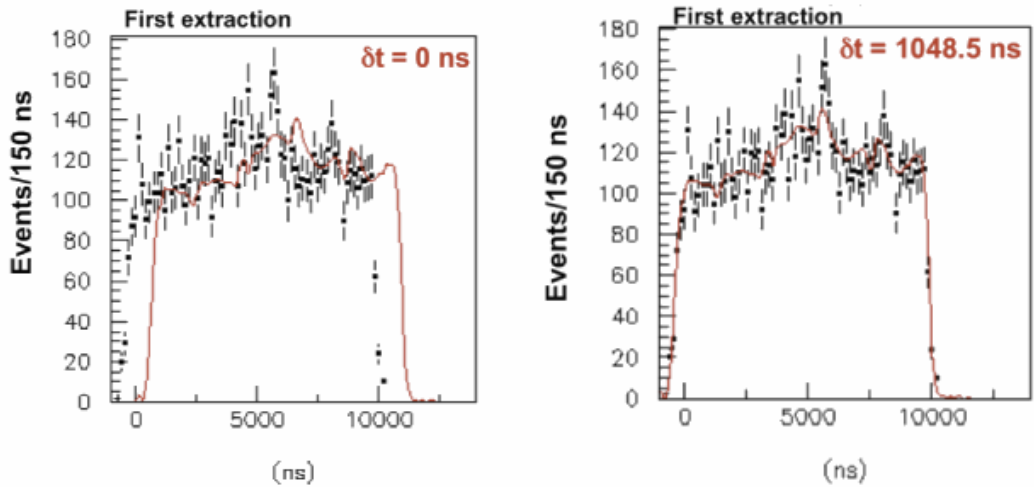}
\caption{The first extraction averaged waveform $\bar{w}_1(t-\Delta)$ (normalized; red curve) and the corresponding data points $\Delta N_{{\rm event}}/\Delta t$ (with $\Delta t = 150 \, {\rm ns}$) given in \cite{Adam} (fig.11). The left figure ($\delta t=0 \, {\rm ns}$) does not contain the shift due to a systematic time bias, the right one ($\delta t = 1048.5 \, {\rm ns}$)  contains both the shift due to the systematic bias ($987 \, {\rm ns}$) and the supplementary delay of $60 \, {\rm ns}$.}
\label{figure2}
\end{figure}

First it must be noticed that if the detected neutrinos at LNGS are time-stamped, their exact creation (in time and position) at CERN is unknown. Therefore it is impossible to measure the individual velocity of each neutrino and only a procedure dealing with the whole pulse is possible. \\
Furthermore the efficiency of the neutrino detection being very low, the number of events detected per unit of time is small,  as a consequence it is not easy to rebuild with precision the instantaneous neutrino rate $\phi_\nu(t) \propto \dfrac{\Delta N_{{\rm event}}}{\Delta t} (t)$. Indeed this procedure requires to count the number of events $\Delta N_{{\rm event}}$ during the interval of time $\Delta t$ and this number $\Delta N_{{\rm event}}$ must be sufficient to have a relative uncertainty (due to statistical effects) $\sqrt{\Delta N_{{\rm event}}}/\Delta N_{{\rm event}}=1/\sqrt{\Delta N_{{\rm event}}}$ small enough. Therefore  $\Delta t$ should not be too small. Of course this introduces an uncertainty on time, or a time resolution, typically $\pm \Delta t/2$, since this procedure gives only access to the averaged rate during $\Delta t$. \\
In fact two approximations of $\bar{\phi}_\nu^{(i)}(t)$, ($i=1,2$) are presented in \cite{Adam}: in fig.11 of \cite{Adam} with $\Delta t =150 \, {\rm ns}$, and in fig.12 of \cite{Adam} with $\Delta t = 50 \,{\rm ns}$. \\
The figure 11 of \cite{Adam} (partially reproduced here in fig.\ref{figure2}) shows the data points of $\bar{\phi}_\nu^{(i)}(t)$ and the waveforms $\bar{w}_i(t-\Delta)$ (normalized), shifted or not by a delay. But this figure might be misleading because:
\begin{itemize}
\item  The graphs labelled with no delay (where $\bar{w}_i(t-\Delta)$ and the $\bar{\phi}_\nu^{(i)}(t)$ data points seem to differ by a shift) are in fact polluted by a (known) systematic bias on time of $987 \, {\rm ns}$,
\item The graphs that perfectly match, labelled with a delay of $1048 \, {\rm ns}$, would have matched as well with the delay of $987 \, {\rm ns}$ corresponding uniquely to the correction of the systematic bias,
\item The difference  $1048-987=61\, {\rm ns}$ is not here significant because the $\phi_\nu$ data points are obtained with the time resolution $\Delta t =150 \, {\rm ns}$.
\end{itemize}
In fact  figure 11 of \cite{Adam} seems to prove that taking into account the systematic bias on time and the averaged knowledge of $\phi_\nu$ (time resolution of $\Delta t =150 \, {\rm ns}$), the usual delayed proportionality rule \eqref{eqn:propor} is compatible with the data (no superluminal effect).\\
To go beyond this result we need a more accurate data analysis. This is done in \cite{Adam} through a {\it likelihood procedure} which is analyzed in the following section. The fig.12 of \cite{Adam} obtained with the time resolution $\Delta t=50 \, {\rm ns}$ is also analyzed below.

\subsubsection{Proving the early arrival of neutrinos}
Before analyzing the procedure used by \cite{Adam}, let us analyze, on a pure logical level, how the data obtained can prove unambiguously the early arrival of neutrinos. \\
Because of the rectangular shape of both pulses $\bar{w}_i(t-\Delta)$ and $\bar{\phi}_\nu^{(i)}(t)$, and because of their long duration ($10 \, \mu {\rm s}$), only the left (leading) edge and the right (trailing) edge of the pulses and their possible shifts contain (possibly) significant informations. But while an early decrease of the neutrino signal (trailing edge) can be possibly due to some unknown property of the proton beam inducing a relation between the proton courant and the neutrino rate different from a simple delayed rule of proportionality, an early beginning of the neutrino signal (leading edge) is the signature of some superluminal effect (at least using arguments of usual Physics). Therefore the only parts of the data that possibly contain an unambiguous information about an early arrival of neutrino are the leading edges of $\bar{w}_i(t-\Delta)$ and $\bar{\phi}_\nu^{(i)}(t)$ (if it is possible to prove that they differ only by a shift). \\
As a consequence, {\bf any numerical procedure built to test the delayed relation \eqref{eqn:propordelay} and that deals with the whole data set contains possibly an hidden {\it logical} bias}: the value of the obtained delay (and its uncertainty) can be in fact polluted  by a deformation of the neutrino signal  not due at all to a superluminal effect.\\
Therefore a possible method is the development of a numerical test that looks for 
\begin{itemize}
\item[(a)] on one hand a possible shift between the leading edges of  $\bar{w}_i(t-\Delta)$ and $\bar{\phi}_\nu^{(i)}(t)$, 
\item[(b)] independently, on the other hand,  a possible shift between the trailing edges of  $\bar{w}_i(t-\Delta)$ and $\bar{\phi}_\nu^{(i)}(t)$.
\end{itemize}
If both computations gives the same result with a good confidence interval, then it can be reasonable to think that the neutrino pulse presents a superluminal effect. If the procedure is unable to give the same delay and/or sufficient confidence intervals (essentially in the leading edge case), then no proof of a superluminal effect exists.\\
\begin{figure}[!ht]
\resizebox{\hsize}{!}{\includegraphics{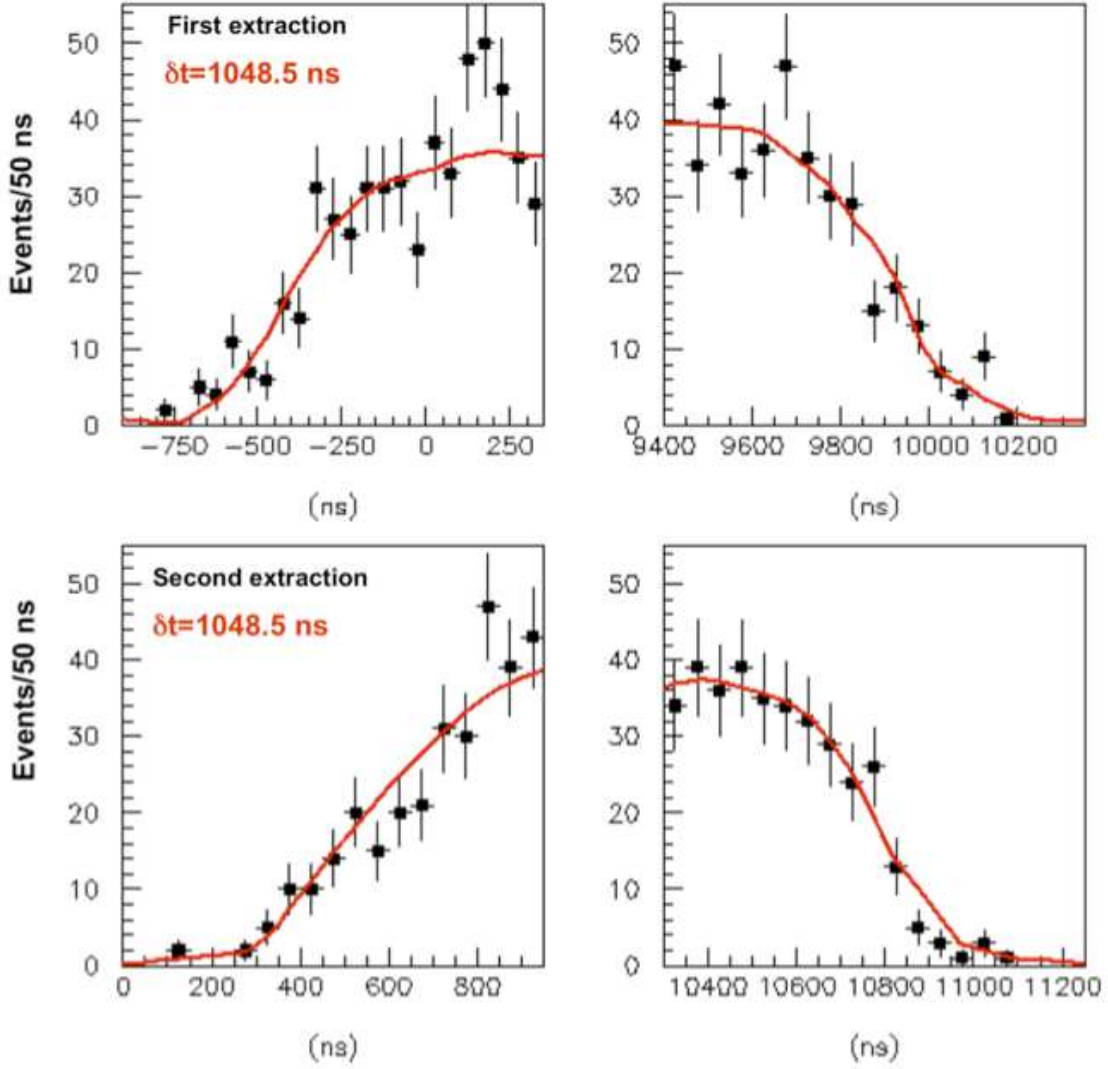}}
\caption{Reproduction of the fig.12 of \cite{Adam} showing the details of the leading and trailing edges of the pulses (waveform in read)  for two successive extractions. The previous delay of $1048.5 \, {\rm ns}$ including $987 \, {\rm ns}$ of systematic bias and $60 \, {\rm ns}$ of supplementary delay is used. The time resolution for the neutrino rate is $\Delta t=50 \, {\rm ns}$.}
\label{figure3}
\end{figure}

The figure 12 of \cite{Adam} reproduced here in fig.\eqref{figure3} shows the behavior of leading and trailing edges of the pulses. The time resolution for computing the neutrino rate was here chosen as $\Delta t=50 \, {\rm ns}$. This introduces an uncertainty on time of about $\pm 25 \, {\rm ns}$, (a little) less than the stated effect of the supplementary delay (advance) of  $60\, {\rm ns}$. \\
Therefore these data are a good starting point to test our ideas about data treatments, even if the full precision of time-stamped events cannot be used in this case and then the expected accuracy is (much) lower. This analysis is done in the following sections.

Let us analyze now the procedure used in \cite{Adam} to obtain OPERA result. It is based on the whole pulse data set, and then it is possibly polluted by some hidden bias as mentioned above. Nevertheless this method is interesting because it allows to get rid of the time resolution problem inherent in the determination of the neutrino rate and then the full precision of time-stamped events can be used. This method uses the fact $\dfrac{d N_{{\rm event}}}{dt} \propto \rho_\nu(t)$ where $\rho_\nu(t)$ is the probability density (in time) of detecting a neutrino and then from eq.\eqref{eqn:propordelay}, $\rho_\nu(t) \propto w(t-\Delta-\tau)$. Assuming the independence of events the probability density $\rho_\nu(t_1,t_2,\dots,t_N)$ for detecting $N$ neutrinos is then
\begin{equation}
\rho_\nu(t_1,t_2,\dots,t_N) \propto \prod_{n=1}^N w(t_n-\Delta-\tau).
\end{equation}
The likelihood method is a bayesian procedure. Given the detection times  $(t_1^{(m)},t_2^{(m)},\dots,t_N^{(m)})$ they attempt to specify the probability law for the parameter $\tau$. Up to a normalization this distribution is the function $\tau \mapsto \rho_\nu(t_1^{(m)},t_2^{(m)},\dots,t_N^{(m)})$ because the function $\rho_\nu(t_1,t_2,\dots,t_N)$ can be seen as a conditional probability. Therefore the most probable value of $\tau$ can be computed as the maximum of this function.\\
In the following section we reanalyze differently the results mentioned in fig.\ref{figure3}.
\section{Our data analysis of the leading and trailing edges}
In this section the waveforms $\bar{w}_i(t)$ are assumed to be normalized and already shifted by  the (known) delay $\Delta=L/c$.
\begin{figure}[!ht]
\resizebox{\hsize}{!}{\includegraphics{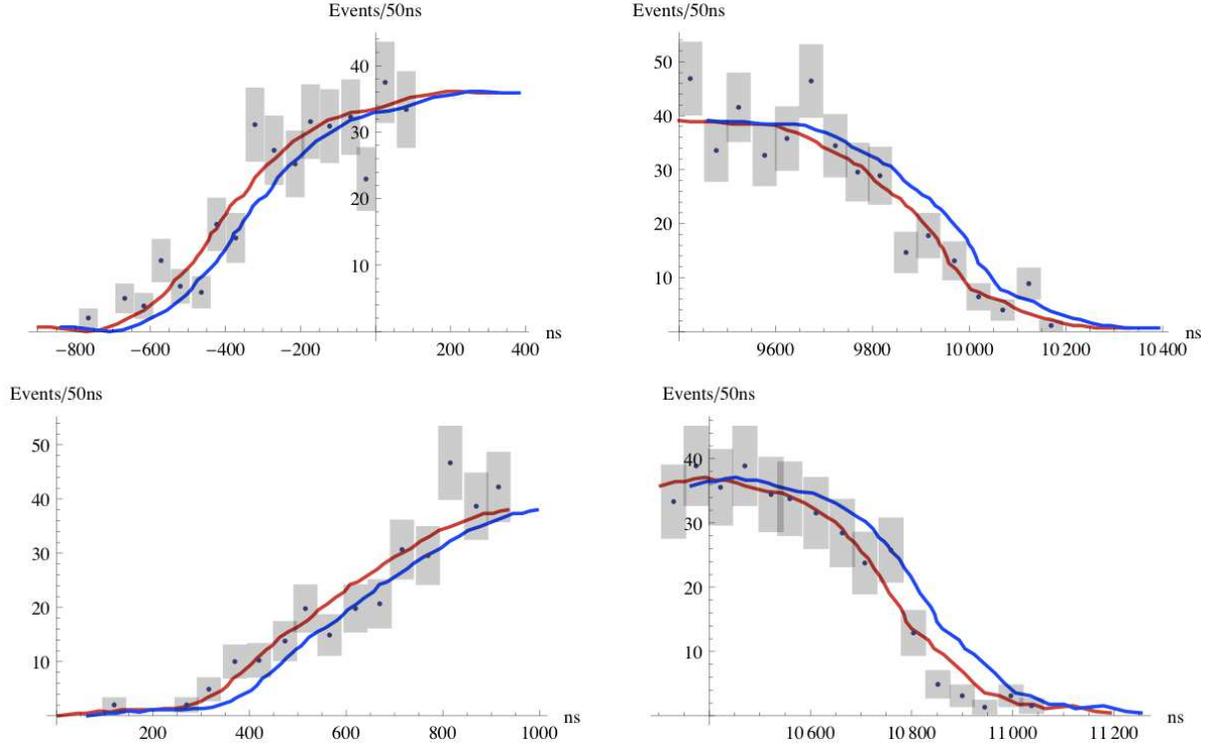}}
\caption{Digitalization of the data of fig.\ref{figure3}. The uncertainties are represented by grey rectangles. The red curves are the delayed waveforms of fig.\eqref{figure3}, while the blue curves are the same waveforms assuming only a travel at the speed of light.}
\label{figure4}
\end{figure}

We digitalized the data of fig.\ref{figure3} to obtain the waveforms $\bar{w}_i(t)$ and the data points of the neutrinos rates $\bar{\phi}_\nu^{(i)}(t)$ (for the leading and trailing edges of the first and second extractions). We also removed the supplementary delay of $60 \, {\rm ns}$ from the digitalized functions to have access to the waveforms $\bar{w}_i(t)$ corrected from the systematic time bias, but without any supplementary shift. The drawings are represented in fig.\ref{figure4}.\\
From these data we developed two types of analysis. The first one is a standard $\chi^2$ method, the second one is more unusual, but very adapted to this situation.
\vspace{0.5cm}

\noindent {\it Remark and Notation}\\
 In the following we remove the index relative to the first and second extractions, and the index $i$ is always relative to the data points of the neutrino rate (points $(t_i,\phi_i)$). 
\subsection{A standard $\chi^2$ method}
\subsubsection{The procedure}
In each case (leading or trailing edge, first or second extraction), {\it assuming the rate model} $\phi(t)=w(t-\tau)$, we can define a first $\chi_0^2(\tau)$ function as 
\begin{equation}
\chi_0^2(\tau)=\dfrac{1}{\Sigma_\phi^2} \sum_{i=1}^N (\phi_i - w(t_i-\tau))^2 \quad {\rm with } \quad \Sigma_\phi^2=\sum_{i=1}^N \sigma_{\phi_i}^2
\end{equation}
where the $\{ \sigma_{\phi_i} \}$ are the (statistical) uncertainties on $\{ \phi_i \}$. \\
Nevertheless this $\chi_0^2(\tau)$ function does not take into account the uncertainty on time induced by the averaged rate. So a better choice is the following $\chi^2(\tau)$ function
\begin{equation}
\label{eqn:chi2}
\chi^2(\tau)=\dfrac{1}{\Sigma_\phi^2} \sum_{i=1}^N \left( \phi_i - \int_{t_i-\delta}^{t_i+\delta} w(t-\tau) \dfrac{dt}{2 \delta} \right)^2 \quad {\rm with} \quad 2\delta=\Delta t=50 \, {\rm ns}
\end{equation}
corresponding to the time averaged rate model $<\phi>(t)=\int_{t-\delta}^{t+\delta} w(u-\tau) \dfrac{du}{2 \delta}$.

Before using the method, we want first to discuss which types of informations are deducible from the $\chi^2$ minimization. By nature this test can only help us to answer to the following question: 
\begin{center}
{\it Assuming the existence of random gaussian noises (with known widths) around the model values, are the data points compatible with this model?}
\end{center}
If the minimization process leads to a minimal value $\chi_{{\rm min}}=\chi(\tau_{{\rm min}}) \le 1$,  our choice of the normalization factor $\Sigma_\phi$ implies that the model for $\tau=\tau_{{\rm min}}$ is effectively compatible with the data. But this is not  sufficient to conclude because:
\begin{itemize}
\item[(a)] Any model function (or value of $\tau$) such that $\chi \le 1$ is also compatible with the data on a statistical point of view; then only the interval $[\tau_1,\tau_2]$ around $\tau_{{\rm min}}$ such that $\chi \le 1$ is meaningful,
\item[(b)] If the uncertainties on $\{\phi_i\}$ are not exactly known the previous interval $[\tau_1,\tau_2]$ can be notably modified,
\item[(c)] If the uncertainties on the data points are large, the condition $\chi \le 1$ can be verified for very different model functions (not reduced to a specific family of one-parameter functional). Then if we are not absolutely certain of the model function type, even a narrow interval $[\tau_1,\tau_2]$ around $\tau_{{\rm min}}$ is not sufficient to validate the model function.
\end{itemize}
With these limitations in mind, we can look at the results.
\subsubsection{The results}
\begin{figure}[!ht]
\resizebox{\hsize}{!}{\includegraphics{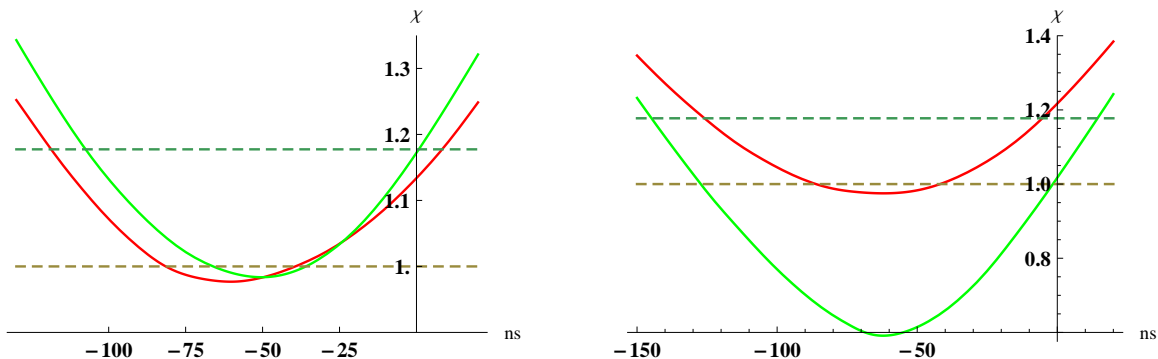}}
\caption{Drawings of $\chi(\tau)$ of eq.\eqref{eqn:chi2}. On the left side:  $\chi(\tau)$ for the leading edges of the two extractions (red curve for the first extraction, green curve for the second one). On the right side:  $\chi(\tau)$ for the trailing edges of the two extractions with the same color codes. The horizontal dashed lines corresponds to the values $\chi=1$ and $\chi=\sqrt{2 \ln2}$.}
\label{figure5}
\end{figure}

\begin{table}[htb!]
  \centering 
\begin{tabular}{|c|c|c|c|c|}
\hline
  & & &  \\
Edge    & $\chi_{\rm min}$ & $\tau_{{\rm min}}$ & $\chi \le 1$& $\chi \le \sqrt{2 \ln2}$  \\
\& Extraction&&&& \\
\hline
\hline
 Leading Edge & & &   \\
 \& First Extraction & $0.97$ & $-60.3 \, {\rm ns}$ & $\Delta \tau=42\, {\rm ns}$&$\Delta \tau=126\, {\rm ns}$ \\
   \hline
  Leading Edge   & & & &  \\
 \& Second Extraction & $0.98$ & $-50.3 \, {\rm ns}$ & $\Delta \tau=30\, {\rm ns}$& $\Delta \tau=108\, {\rm ns}$ \\
   \hline
 Trailing Edge   & & & &  \\
 \& First Extraction & $0.97$ & $-61.9 \, {\rm ns}$ & $\Delta \tau=44\, {\rm ns}$ &$\Delta \tau=120\, {\rm ns}$ \\
 \hline
  Trailing Edge  & & &  & \\
  \& Second Extraction & $0.59$ & $-62.2 \, {\rm ns}$ &$\Delta \tau=125\, {\rm ns}$& $\Delta \tau=160\, {\rm ns}$ \\
 \hline
\end{tabular}
\caption{Summary of the $\chi^2$ minimization corresponding to the fig.\ref{figure5}.}
\label{table1}
\end{table}

The drawings of $\chi(\tau)$ for the different leading and trailing edges (for the two extractions) are represented in fig.\ref{figure5}. \\
The parameters deduced from the minimization are summarized on the table \ref{table1}.
\begin{itemize}
\item The values $(\tau_{{\rm min}},\chi_{{\rm min}})$ represent the values obtained from the minimization (for the four edges). All $\chi_{{\rm min}}$ verify the required condition $\chi_{{\rm min}} \le 1$.
\item We have also computed the widths $\Delta \tau$ of the interval $[\tau_1,\tau_2]$ such that $\chi \le 1$ and $\chi \le \sqrt{2 \ln2}$, the value $\sqrt{2 \ln2}$ coming from the evaluation of uncertainties from the width at half-height (for a gaussian) rather from its parameter $\sigma$. Each value of $\Delta \tau$ can give an estimate of the uncertainty $\delta \tau=\Delta \tau/2$ on the value of $\tau$. In the present case the obtained values are very different.
\end{itemize}

\subsubsection{Discussion}
 At first sight of  minima $\tau_{{\rm min}}$ in table \ref{table1}, this procedure seems to confirm the model function and the stated result in \cite{Adam} of an early arrival of neutrinos characterized by the delay $\tau \simeq - 60\, {\rm ns}$. Indeed the four values of $\tau_{{\rm min}}$, computed independently, agree with this value.  But the width $\Delta \tau$ of the intervals such that $\chi \le 1$ ($\Delta \tau \simeq 40 {\rm ns}$) and $\chi \le \sqrt{2 \ln2}$ ($\Delta \tau \simeq 120 {\rm ns}$) gives a completely different picture: because of the high sensitivity to the data uncertainties, $\delta \tau=\Delta \tau/2$ is possibly of the same order of magnitude as the sought shift. Therefore, even if we can suspect a real shift, it is impossible to affirm that it exists. This paradox cannot be solved on this level: it requires the development of a new tool. This is done in the next section.

\subsection{Testing the type of the model function}
In this section we develop a statistical method that allows to test directly the validity of the previous model function $<\phi>(t)=\int_{t-\delta}^{t+\delta} w(u-\tau) \dfrac{du}{2 \delta}$, rather than looking uniquely for the best value of $\tau$.

\subsubsection{The procedure}
It is a matter of fact that in the case of edges the function $w(t)$ is either increasing (leading edge), either decreasing (trailing edge). This property is unchanged using our averaged function 
\begin{equation}
\hat{w}(t)=\int_{t-\delta}^{t+\delta} w(u) \dfrac{du}{2 \delta}.
\end{equation}
We deduce that this function is invertible (in the edge domains of the pulses). Then if the rate verifies $\phi(t)=\hat{w}(t-\tau)$, we deduce
\begin{equation}
 \hat{w}^{-1}(\phi(t))=t-\tau
 \end{equation}
 We deduce that our data points $\{(t_i,\phi_i)\}_{i=1}^N$ must verify
\begin{equation}
\tau=t_i-\hat{w}^{-1}(\phi_i)
\end{equation}
Therefore we have only to verify that the set of data $\{\tau_i=t_i-\hat{w}^{-1}(\phi_i)\}_{i=1}^N$ is compatible with a unique constant. Verifying that a cloud of points is reasonably in the neighborhood of an horizontal line is very easy. Then independently of the value of $\tau$, this test allows to verify if the model function is reasonable, because each experimental point generates a possible value of $\tau$. Furthermore all usual statistical tests can be tried on the cloud of points.\\
To keep the time structure of the data we built the new set of time-indexed data points $\{(t_i, \tau_i=t_i-\hat{w}^{-1}(\phi_i)) \}_{i=1}^N$ for the four sets (edges) of experimental results analyzed in the previous section. 

\subsubsection{The results}
\begin{figure}[!ht]
\resizebox{\hsize}{!}{\includegraphics{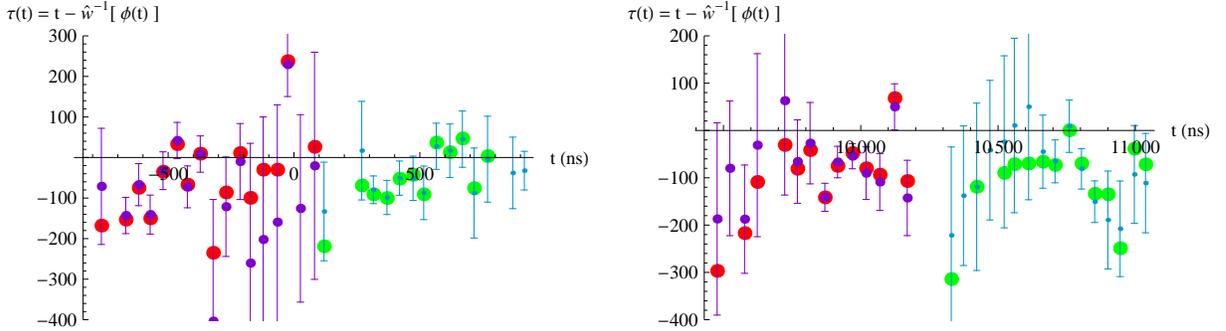}}
\caption{Experimental points of fig.\ref{figure3} and fig.\ref{figure4} transformed into the sets $\{(t_i, \tau_i=t_i-\hat{w}^{-1}(\phi_i)\}_{i=1}^N$. On the left side: points from the leading edges (red: first extraction, green: second extraction). On the right side: points from the trailing edges, same color codes.}
\label{figure6}
\end{figure}

The results are illustrated in  fig.\ref{figure6}. \\
Let us remark that  some $\phi$ data points are outside the range of the function $\hat{w}$, so they cannot have a reciprocal image by $\hat{w}^{-1}$. In the same way the error bars in the $\phi$ domain are not always contained in the range of $\hat{w}$, so taking the reciprocal image by $\hat{w}^{-1}$, they are not anymore centered on the experimental points.\\
A simple glance at fig.\ref{figure6} shows that the points of the leading edges do not seem to be aligned on a horizontal line that differs from the $t$-axis, whereas this condition is essential to conclude positively to a superluminal effect. 

\begin{table}[htb!]
  \centering 
\begin{tabular}{|c|c|c|c|cl}
\hline
  & & &  \\
Edges    & $<\tau>$ & $<\tau^2>$ & $\sigma_\tau$  \\
& & & \\
\hline
\hline
 & & &    \\
Leading Edges & $-51.8\, {\rm ns}$ & $11124.7 \, {\rm ns}^2$ & $91.8\, {\rm ns}$ \\
 \hline
  & & &    \\
Trailing Edges & $-101.9\, {\rm ns}$ & $17233.3 \, {\rm ns}^2$ & $82.7\, {\rm ns}$ \\
 \hline
\end{tabular}
\caption{Summary of the statistical computations on $\tau$ from the data points of fig.\ref{figure6}.}
\label{table2}
\end{table}

Since we have now a sequence of values of $\tau$, we can compute the mean value and the standard error (or uncertainty) in a usual statistical sense. The calculations have been made independently on the leading edges and on the trailing edges: the results are summarized on the table \ref{table2}.
\newpage

\subsubsection{Discussion}
It can be divided in two parts:
\begin{itemize}
\item[(a)] Leading edges: the high value of the ratio $\dfrac{\sigma_\tau}{|<\tau>|}
\simeq 1.8$ implies that it is impossible to define a reliable non-zero value of $\tau$ because of the too noisy data. Then the superluminal effect is not unambiguously validated and if we are conservative, the value $\tau \simeq 0$ remains the most reasonable choice.
\item[(b)] Trailing edges: the value of the ratio $\dfrac{\sigma_\tau}{|<\tau>|}\simeq 0.8$ is compatible with a possible noisy real shift.
\end{itemize}

\begin{figure}[!ht]
\resizebox{\hsize}{!}{\includegraphics{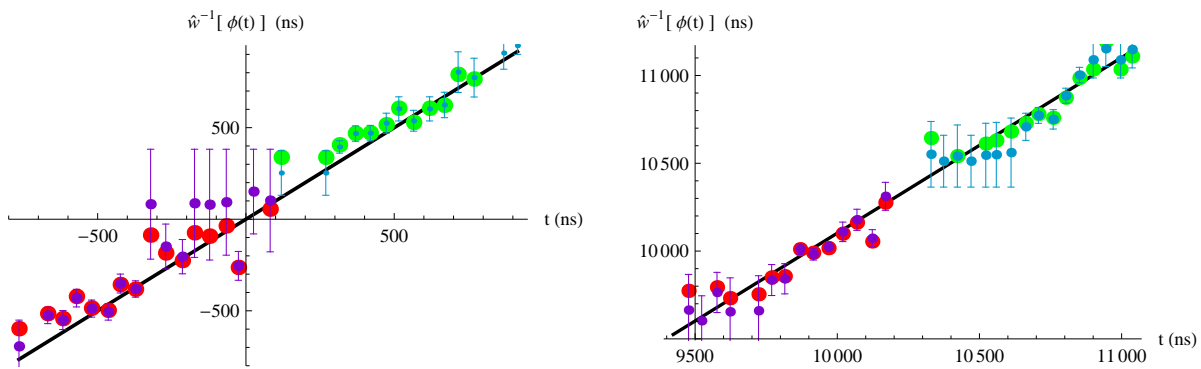}}
\caption{Experimental points of fig.\ref{figure3} and fig.\ref{figure4} transformed into the sets $\{( \, t_i, \hat{w}^{-1}(\phi_i)\, )\}_{i=1}^N$. On the left side: points from the leading edges (red: first extraction, green: second extraction) and the fitting function $f(t)=t$. On the right side: points from the trailing edges (same color codes) and the fitting function $f(t)=t+102$.}
\label{figure7}
\end{figure}

\noindent These results can be also (qualitatively) visualized plotting the data set $\{ (\, t_i,\hat{w}^{-1}(\phi_i)\,) \}$, since according to the model $\hat{w}^{-1}(\phi(t))=t-\tau$. The resulting plot is given in the figure \eqref{figure7}. The points coming from the leading edges fit well the curve $f(t)=t$, while the points of the trailing edges fit well the curve $f(t)=t+102$. This representation is also natural since points coming from successive extractions can be represented on the same time axis and compared with the same function $f(t)$.

Put together these results show {\bf at the very least} that the experimental data do not permit to (completely) exclude one of the following solutions:
\begin{itemize}
\item a very noisy (leading edge) early arrival of neutrinos,
\item a small deformation of the neutrino signal (with respect to the proton pulse) in the trailing edge,
\item a mixing of the previous solutions.
\end{itemize}
Therefore, as long as the assumption of a neutrino pulse deformation cannot be ruled out, it is impossible to claim  that ``{\it the measurement indicates an early arrival time of CNGS muon neutrinos with respect to the one computed assuming the speed of light in vacuum}''.

Let us finish this discussion by a remark on the possible benefit of a Monte Carlo simulation (followed by our data analysis). In fact no additional information can be obtained, regardless of the simulation type, if no additional experimental data are given (for example the distribution of protons energy in the trailing edge). Indeed, since the true physical phenomena involved are unspecified, and since the observed data are (at least) ambiguous, no simulation (with a good or bad agreement with the data) is able to exclude or confirm assumptions. If the agreement is bad, this  can be due to a partial description of the true physical phenomena, but {\it \`a priori} this does not exclude anything (mixed effect). If the agreement is good, since the data are ambiguous, this is not enough to prove that other physical assumptions would not give the same results.

\section{Conclusion}
The OPERA experiment is very impressive due to the very high precision of the data, the latter being obtained from a huge and elaborate instrumental setup. Nevertheless the claim of a superluminal propagation of the neutrino beam cannot be deduced {\it unambiguously} from their data. The reasons lie in: (a) the smallness of the sought effect, (b) the experimental uncertainties, (c) a special feature of the data set. Put together these circumstances lead easily to wrong conclusions. In fact our data analysis shows that the time distribution of the neutrino packet is {\it more probably} not a time translation of the proton one, but rather a slight deformation (trailing edge). At the very least this solution cannot be ruled out.

\end{document}